# ODT Flow Explorer: Extract, Query, and Visualize Human Mobility

Zhenlong Li[1*], Xiao Huang[2], Xinyue Ye[3], Xiaoming Li[4]

[1]*Geoinformation and Big Data Research Lab, Department of Geography, University of South Carolina, Columbia, SC, USA*

[2]*Department of Geosciences, University of Arkansas, Fayetteville, AR, USA*

[3]*Department of Landscape Architecture & Urban Planning, Texas A&M University, TX, USA*

[4]*Department of Health Promotion, Education, and Behavior, University of South Carolina, Columbia, SC, USA*

[*]Email: zhenlong@sc.edu

## Abstract

Understanding human mobility dynamics among places provides fundamental knowledge regarding their interactive gravity, benefiting a wide range of applications in need of prior knowledge in human spatial interactions. The ongoing COVID-19 pandemic uniquely highlights the need for monitoring and measuring fine-scale human spatial interactions. In response to the soaring needs of human mobility data under the pandemic, we developed an interactive geospatial web portal by extracting worldwide daily population flows from billions of geotagged tweets and United States (U.S.) population flows from SafeGraph mobility data. The web portal is named ODT (Origin-Destination-Time) Flow Explorer. At the core of the explorer is an ODT data cube coupled with a big data computing cluster to efficiently manage, query, and aggregate billions of OD flows at different spatial and temporal scales. Although the explorer is still in its early developing stage, the rapidly generated mobility flow data can benefit a wide range of domains that need timely access to the fine-grained human mobility records. The ODT Flow Explorer can be accessed via http://gis.cas.sc.edu/GeoAnalytics/od.html.

## 1. Introduction

Prediction and control of the spread of infectious diseases such as COVID-19 benefit greatly from our growing computing capacity to quantify fine-scale human movement (Hancock et al., 2014; Kraemer et al., 2020). In response to the soaring needs of human mobility data during the COVID-19 pandemic, we extracted the worldwide daily population flows from billions of geotagged tweets and SafeGraph data, and developed an interactive geospatial web portal, called ODT (Origin-Destination-Time) Flow Explorer (http://gis.cas.sc.edu/GeoAnalytics/od.html, Figure 1), that allows researchers to query, aggregate, visualize, and download daily human movement data at various geographic scales. This article briefly explains how we extracted of population movement from Twitter and SafeGraph data, demonstrates how the ODT Flow Explorer can be used to query, visualize, and download human mobility data, and discusses the limitations of each dataset.

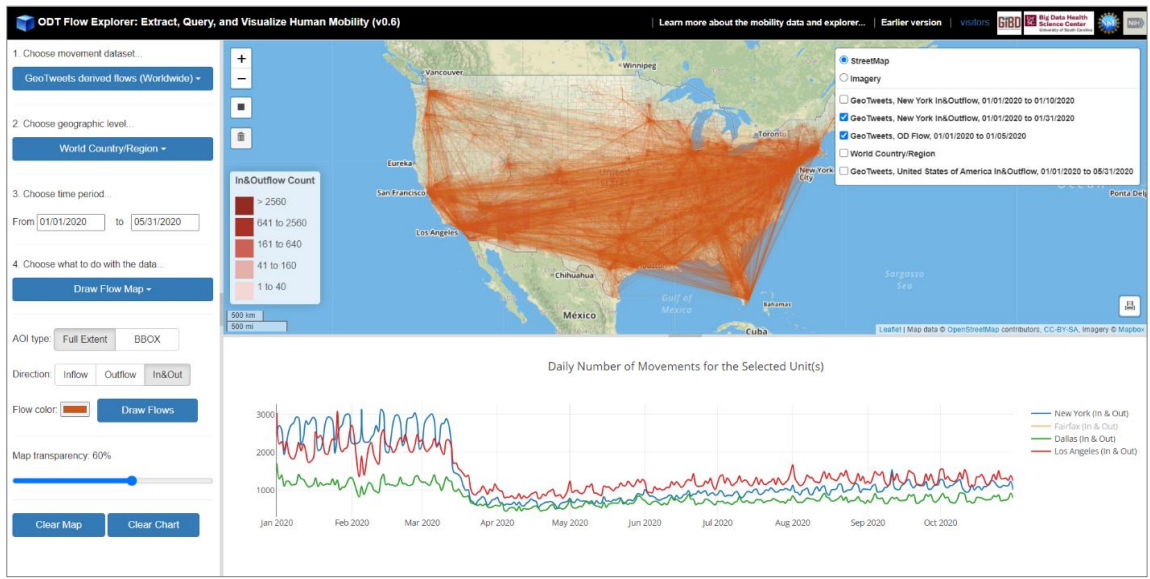

Figure 1. User Interface of ODT Flow Explorer (v0.66).

## 2. Extracting Human Mobility

We derived the human movement data in the format of origin-destination (OD) flows from two data sources: Worldwide geotagged tweets collected using the Twitter public API (https://developer.twitter.com/en/docs/twitter-api); U.S. mobile device-based Social Distancing Metrics provided by SafeGraph (https://docs.safegraph.com/docs/social-distancing-metrics).

### *2.1 Extracting Daily OD Flows from Geotagged Tweets*

The daily OD flows derived from geotagged tweets are the combination of Twitter users' single-day movement and cross-day movement. The concept of single-day and cross-day movements was introduced in Huang et al. (2020). In general, the single-day movement represents the users' daily maximum travel distance of all locations relative to the initial location, and cross-day movement measures the mean center shift between two consecutive days. Following Martin et al. (2020), we removed the non-human tweets (tweets posted by bots, such as weather reports and job offers) by checking the tweet source. For example, tweets automatically posted for job offers from the source TweetMyJOBS were removed. We also excluded the tweets that are geotagged with spatial resolution coarser than the city level. After the data cleaning, we derived 2.1 billion (2,148,780,155) geotagged tweets posted by over 21 million (21,777,336) Twitter users from 01/01/2019 to 10/31/2020. Following the single-day and cross-day approach, we further extracted over 591 million (591,417,926) user-level daily OD flows covering the whole world. The process was performed using Apache Hive (https://hive.apache.org) coupled with Esri GIS tools for Hadoop (http://esri.github.io/gis-tools-for-hadoop) on our Hadoop computing environment. Note that the Twitter-derived OD flows do not consider users' home location. The movements were directly derived from the locations of geotagged tweets at the Twitter user level on a daily basis.

### *2.2 Extracting Daily OD Flows from SafeGraph Data*

We extracted the daily OD flows in the U.S. using Social Distancing Metrics (SDM) data downloaded from SafeGraph. There are 23 fields in the SDM table, and we used 3 of them to derive the population movement, including *origin_census_block_group*, *destination_cbgs*, and *date_range_start*. The *origin_census_block_group* is the unique 12-digit FIPS code for the Census Block Group. *destination_cbgs* contains a list of key-value pairs with the key indicating the destination census block group (from the origin census block group) and "value is the number of devices with a home in *census_block_group* that stopped in the given destination census block group for >1 minute during the time period" (https://docs.safegraph.com/docs/social-distancing-metrics). The *date_range_start* was used to extract the date information. Based on the three fields, we generated an OD table with each row showing the number of devices from an original block group to a destination block group on a specific day. The new OD table contains over 6 billion (6,144,802,397) block group level daily OD flows for 2019 and over 3.7 billion (3,770,910,837) daily OD flows for 2020 (updated to Sep. 30) covering the U.S. The process was performed in Apache Hive environment (the HiveQL used to generate the OD table is provided at the end of the document). Note that the SafeGraph-derived OD flows consider devices' home location (the movements are originated from home). For example, a flow of 100 devices (users) from county A to county B indicates that the home location of the 100 devices is in county A.

*2.3 Aggregating Daily OD Flows at various geographic levels*

We further aggregated the billions of daily OD flows (Twitter-derived flows at the user level for the whole world and SafeGraph-derived flows at the census block group level for the U.S.) to various geographic scales, including Countries (Twitter only), Worldwide first-level country subdivisions (Twitter only), U.S. States (Twitter and SafeGraph), U.S. Counties (Twitter and SafeGraph), and U.S. Census Tracts (SafeGraph). The spatially aggregated daily OD flows are available in the tool for exploration and download.

## 3. Origin-Destination-Time (ODT) Data Cube

To efficiently manage, query, and aggregate billions of OD flows at different spatial and temporal scales, we developed an Origin-Destination-Time data cube (ODT cube) as a conceptual data model for the ODT Flow Explorer (Figure 2). In the ODT data cube, origin (O) and destination (D) are a set of places or locations (e.g., administrative boundaries such as county, state, and country, or grids) that can be displayed with a map. Each cell in the data cube has a value that indicates the number of flows from the origin location to the destination location during a specific time period (e.g., in an hour, a day, or a month). Three types of matrices can be derived from the ODT data cube: origin-destination (OD) matrix quantifies the population flows between all the origin and destination locations during a time period. Destination-time (DT) matrix captures the number of incoming flows to all destination locations from a specific origin location over a series of times. Similarly, an origin-time (OT) matrix captures the number of outgoing flows from all origins to a specific destination over a series of times. The ODT Flow Explorer aims to provide an interactive interface for on-the-fly querying, slicing, aggregating, and visualizing the ODT data cube. Backed by a high-performance computing cluster, the queries generally take less than 15 seconds in our computing environment.

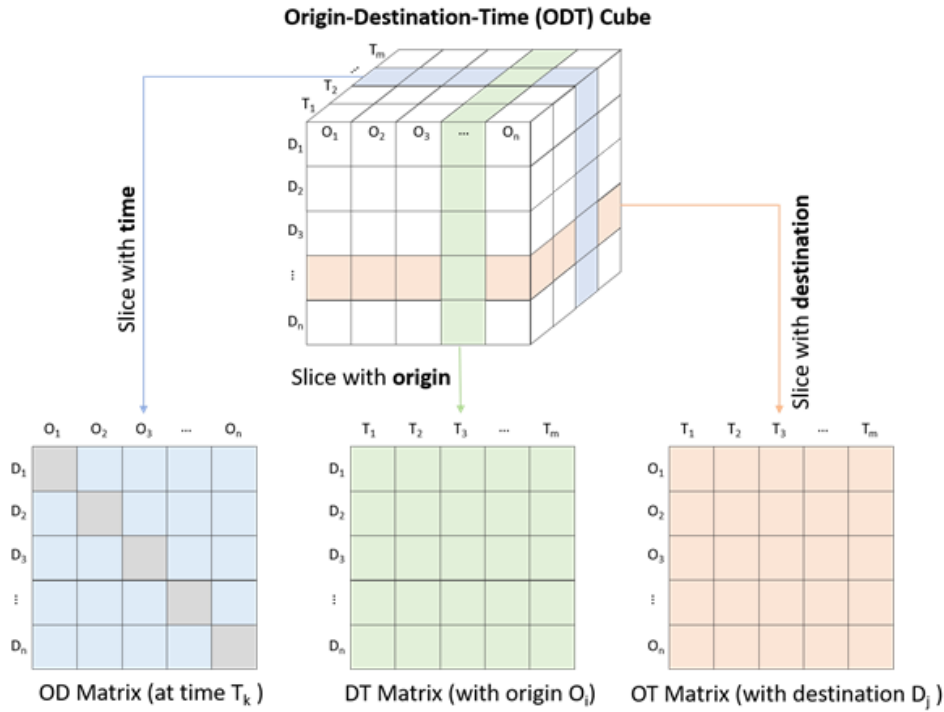

Figure 2. Illustration of Origin-Destination-Time Cube for big OD data query analytics.

## 4. Demonstration

The ODT Flow Explorer allows researchers to explore daily population mobility aggregated at various geographic levels and scales with a few clicks. Most of the components and buttons of the Explorer are self-explanatory (tooltips are also available). Below are the general steps to use this tool: (1) Users start by choosing a movement dataset they are interested in. Currently, the available selections are Twitter derived flows and SafeGraph derived flows; (2) choose a geographic level from the following list: U.S. County (available for both Twitter and SafeGraph), U.S. State (Twitter and SafeGraph), Worldwide Country/Region (Twitter). U.S. Census Tract (SafeGraph) and worldwide first-level country subdivisions from Twitter will be added in the future; (3) the next step is to choose a time period. Currently, only 2020 data are included. Twitter-derived mobility data are updated to October 31, 2020 and SafeGraph-derived mobility data are updated to September 30, 2020; (4) once the dataset, geographic level, and time period are selected, users choose what to do with the selected data (a subset of the ODT data cube). Four options are available (as of version 0.6): *Choropleth Map*, *Flow Map*, *Daily Cross-unit Movements*, and *Download*.

For the ***Choropleth Map*** option, users click on the map to select a geographic unit such as a county or a state to display its aggregated movement between other units as a choropleth map. Flow directions (*Inflow*, *Outflow*, and *In & Out*) can be configured. *Inflow* refers to the number of users/devices from other units moving to the selected unit during the selected time period. *Outflow* refers to the number of users/devices moving from the selected unit to other units. *In & Out* contains the movements from both directions. Figure 3 left shows SafeGraph-derived county population flows to New York County (Manhattan) from 03/08/2020 to 03/14/2020. The right map shows the flows to New York County for the following week (03/15/2020 to 03/21/2020). Figure 4 left shows state-level population flows from/to South Carolina on 01/01/2020, and the right map shows the country level movement.

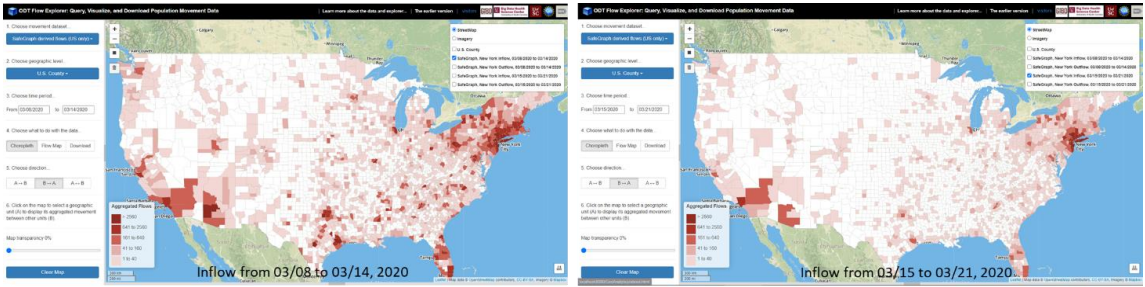

Figure 3. Left: SafeGraph-derived county population flows to New York County from 03/08/2020 to 03/14/2020. Right: population flows for the following week (03/15/2020 to 03/21/2020).

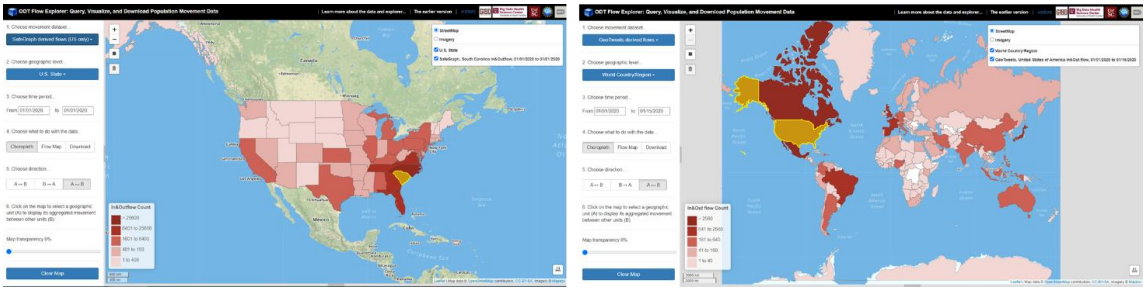

Figure 4. Left: State-level population flows from/to South Carolina on 01/01/2020 based on SafeGraph, Right: Twitter-derived country-level movement from/to the U.S. from 01/01/2020 to 01/15/2020.

For the *Flow Map* option, origin-destination flow lines are directly drawn on the map based on the selected dataset, geographic level, and time period. Users can choose the area of interest (AOI) by drawing a bounding box on the map or use the full spatial coverage of the data. Flow direction (*Inflow*, *Outflow*, and *In & Out*) and the flow color can also be configured. The width of each flow is weighted based on the number device/user movements for display only. Figure 5 shows county-level population movement from 01/01/2020 to 01/05/2020 based on Twitter (left) and SafeGraph (right). Note that for SafeGraph-derived mobility, only flows with aggregated device number great than 20 within the selected time period are displayed to make the number of returned flows manageable for the visualization. WebGL-enabled mapping components such as kepler.gl will be integrated in later versions to overcome this limitation.

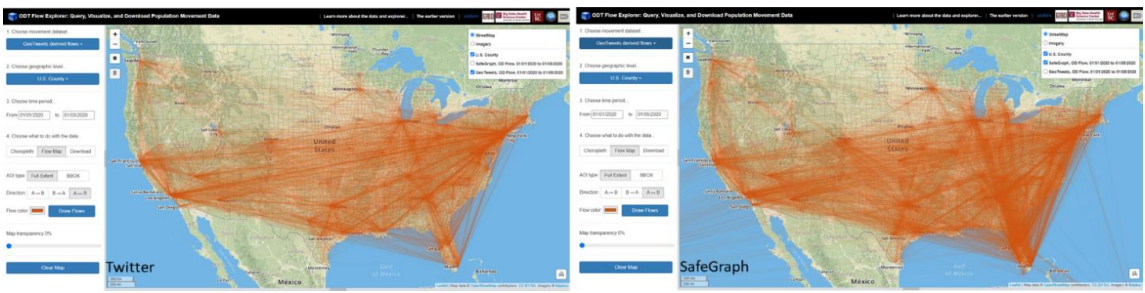

Figure 5. County-level population flows from 01/01/2020 to 01/05/2020 derived from Twitter (left) and SafeGraph (right).

For the *Daily Cross-unit Movements* option, the daily number of movements for a selected geographic unit (e.g., county or country) and year is computed and displayed as a time series chart. The operation is performed by querying the origin-time (OT) matrix and destination-time (DT) matrix. The direction option has four selections: *Inflow*, *Outflow*, *In&Out*, and *Intraflow*.

*Inflow* refers to the number of daily users/devices from all the other units moving to the selected unit. *Outflow* refers to the number of daily users/devices moving from the selected unit to all the other units. *In&Out* contains the daily movements from both directions. *Intraflow* refers to the number of daily movements within the selected unit (flows with a movement distance greater than zero but not crossing the unit boundary). Figure 6 shows the county-level daily outflow movements for New York County and Los Angeles County in the U.S. from 01/01/2020 to 09/30/2020 (based on SafeGraph-derived OD data). Figure 7 shows the country level daily intraflow movements for France, Spain, and Argentinafrom 01/01/2019 to 10/31/2020 (based on Twitter-derived OD data). The impact of the COVID-19 pandemic on human mobility is well reflected from the charts at different geographic scales.

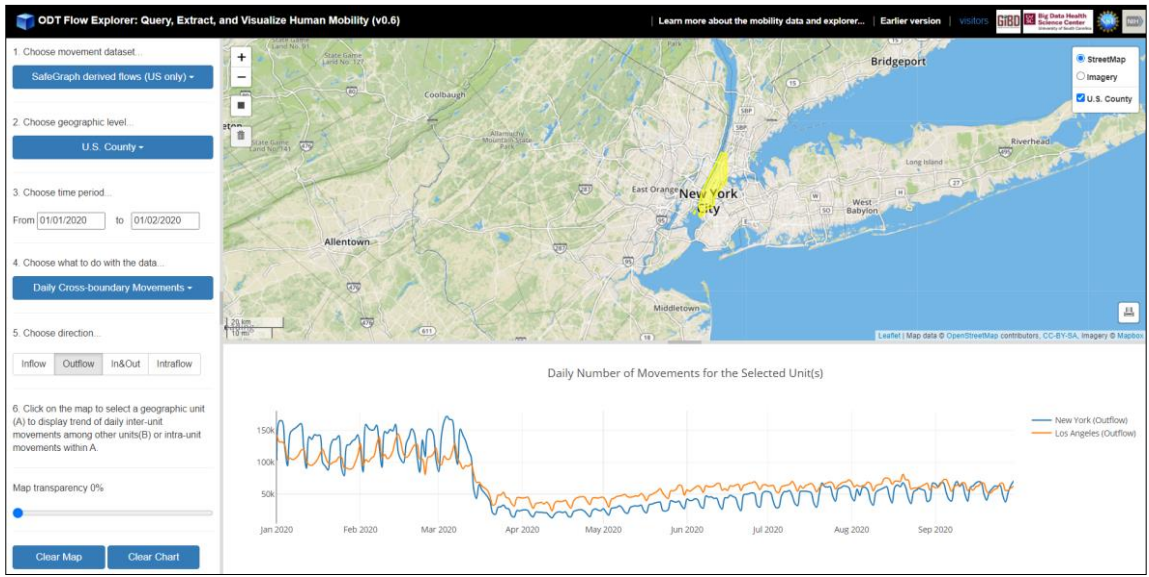

Figure 6. Daily number of county-level outflow movements for New York County and Los Angeles County in the U.S. from 01/01/2020 to 09/30/2020 (based on SafeGraph-derived flows).

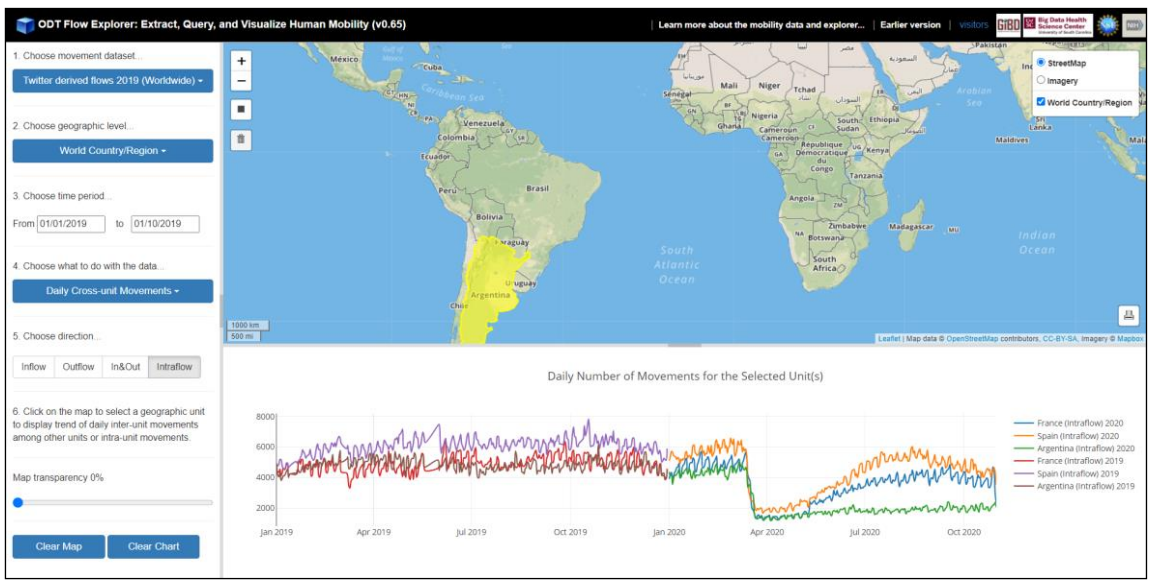

Figure 7. Daily number of country-level intraflow movements for France and Spain from 01/01/2019 to 10/31/2020 (based on Twitter-derived flows).

Figure 8 shows the daily number of intraflow movement (left figure) and inflow/outflow movement (right figure) for Japan from 01/01/2020 to 10/31/2020 based on Twitter-derived flows. The intraflow movement reveals human mobility dynamics within the country in responding to the pandemic. The in&out flow movement, on the other hand, shows the international travels for Japan started to decrease in early March, 2020 and stay on a low level since then.

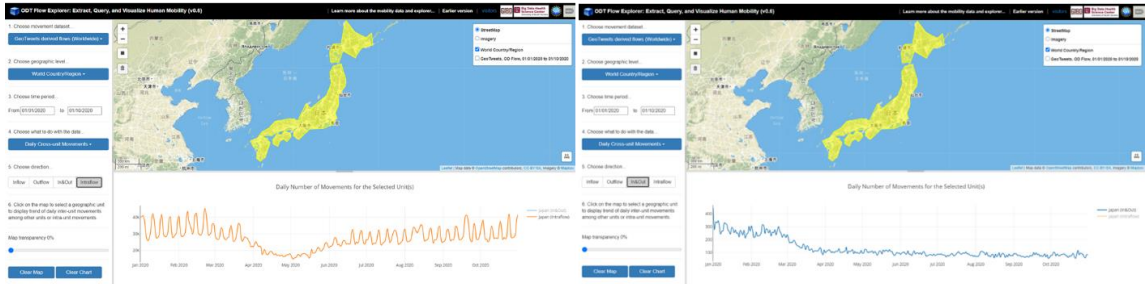

Figure 8. Daily number of intraflow movement (left) and in & out flow movement (right) for Japan from 01/01/2020 to 10/31/2020 (based on Twitter-derived flows).

Lastly but most importantly, users can **extract and download the mobility data** (mobility matrix) by selecting their interested dataset, geographic levels, geographic area, time period, and aggregation type as CSV (comma-separated values) files for further analysis or integrating with predictive models. Figure 9 shows over 2.7 million county-level daily flows were extracted and downloaded for the selected area (flows that are from/to the bbox). Each row in the CSV file contains origin county (*o_fips*), destination county (*d_fips*), date (year, month, day), number of devices/users moved from origin to destination (*cnt*), and mean center of all flow origins (*o_lat, o_lon*) and flow destinations (*d_lat, d_lon*). If the *Aggregated* option is selected, the data file becomes a mobility matrix with the summed number of devices/users moved from origins to destinations during the selected time period.

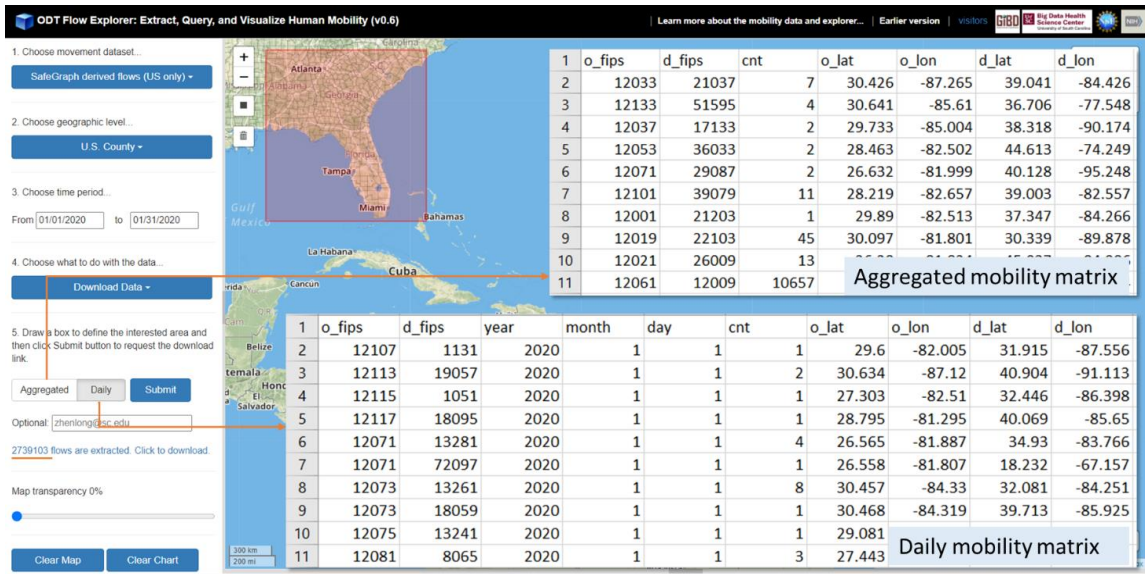

Figure 9. Daily and aggregated mobility matrices were downloaded as CSV files.

## 5. Data Limitations

*Twitter-derived population flows:* Twitter data have intrinsic limitations which have been examined by a number of studies (e.g., Li et al., 2013; Malik et al., 2015; Jiang et al., 2019). Twitter is not proportionally used by different population groups and thus shows demographic and socioeconomic biases. In addition, geotagged tweets collected from the free public Twitter API (about 1% of the whole Twitter stream) are sparse and not enough to capture the temporal patterns at the daily level for less populated areas. This is particularly the case when deriving county level daily population flows because a Twitter user was included only when the user posted at least two tweets on a day or posted tweets on at least two consecutive days. Another limitation is that the dynamics of people's Twitting activities (e.g., people tend to tweet more during big events) as well as the changing of Twitter's internal API affect the daily number of tweets being collected. Studies using the Twitter-derived flow data should be aware of these limitations when interpreting results and reaching conclusions.

*SafeGraph-derived population flows:* SafeGraph data have a high penetration rate (~10% of mobile devices in the U.S.) and well represent the U.S. population groups according to SafeGraph (2019). So the flows derived from SafeGraph are more condensed than Twitter-derived flows, which overcomes the Twitter data limitations. One downside of the SafeGraph-derived mobility data, comparing to Twitter data, is that the data are only freely available in the U.S. dating back to 2019. By enabling intuitive exploration and comparison of the two mobility datasets in the ODT Flow Explorer, this work highlights the importance and necessity of sharing and fusing multiple data sources for human mobility studies.

## 6. Future Development

The ODT Flow Explore is still in its early stage. As the next step, we will add mobility data aggregated for other geographic levels, including the U.S. census tracts (SafeGraph) and worldwide first-level country subdivisions (Twitter). We will also periodically update the mobility datasets as new data (Twitter and SafeGraph) become available. From the function perspective, we plan to add WebGL support (such as kepler.gl) to the system so that it can handle large datasets visualization more efficiently. Currently, the flow visualization functions are very basic and become slow when a large number of records are returned from the query.

## Appendix

HiveQL was used for extracting the census block group level OD flows (table: sg_od) from SafeGraph Social Distancing Metrics dataset (table: sg_social_distancing). The queries need to run in Hive on a Hadoop environment (https://hive.apache.org):

*create view destination_list as*
*select origin_census_block_group, year, month, day,*
*split(translate(substr(destination_cbgs, 2, length(destination_cbgs) - 2),"\"",""), ",") as destinations*

*from sg_social_distancing;*
*create table sg_od as*
*select origin_census_block_group as origin_bg, split(blck,":")[0] as destination_bg, cast(split(blck,":")[1] as int) as device_count, year, month, day from destination_list*
*LATERAL VIEW explode(destinations) dest_table as blck;*

**Acknowledgment:** The study and system development were supported by National Science Foundation (NSF) under grant 2028791, the National Institute of Allergy and Infectious Diseases (NIAID) of the National Institutes of Health (NIH) under grant R01AI127203-4S1, and the University of South Carolina COVID-19 Internal Funding Initiative under grant 135400-20-54176. The funders had no role in study design, data collection and analysis, or preparation of this article/system.

**References**

Hancock PA, Rehman Y, Hall IM, Edeghere O, Danon L, House TA, Keeling MJ. (2014). Strategies for controlling non-transmissible infection outbreaks using a large human movement data set. PLOS Comput Biol Public Library of Science; 2014;10(9):e1003809.

Kraemer, M. U., Yang, C. H., Gutierrez, B., Wu, C. H., Klein, B., Pigott, D. M., ... & Brownstein, J. S. (2020). The effect of human mobility and control measures on the COVID-19 epidemic in China. Science, 368(6490), 493-497.

Huang X., Li Z., Jiang Y., Li X., Porter D., (2020) Twitter reveals human mobility dynamics during the COVID-19 pandemic, PloS One, https://doi.org/10.1371/journal.pone.0241957

Martín, Y., Cutter, S. L., Li, Z., Emrich, C. T., & Mitchell, J. T. (2020). Using geotagged tweets to track population movements to and from Puerto Rico after Hurricane Maria. Population and Environment, 1-24.

Li, L., Goodchild, M. F., & Xu, B. (2013). Spatial, temporal, and socioeconomic patterns in the use of Twitter and Flickr. Cartography and geographic information science, 40(2), 61-77.

Malik, M. M., Lamba, H., Nakos, C., & Pfeffer, J. (2015). Population bias in geotagged tweets. People, 1(3,759.710), 3-759.

Jiang, Y., Li, Z., & Ye, X. (2019). Understanding demographic and socioeconomic biases of geotagged Twitter users at the county level. Cartography and geographic information science, 46(3), 228-242.

SafeGraph, What about bias in the SafeGraph dataset? (2019), https://www.safegraph.com/blog/what-about-bias-in-the-safegraph-dataset, last accessed on November 8, 2020